\documentclass[pre,reprint,superscriptaddress]{revtex4-1}
\usepackage{amssymb,amsmath}
\usepackage{graphicx}
\newcommand{\ignore}[1]{{}}

\begin{document}
\title{Isolation probabilities in dynamic soft random geometric graphs}
\author{Carl P. Dettmann}
\affiliation{School of Mathematics, University of Bristol, University Walk, Bristol BS81TW, UK}
\author{Orestis Georgiou}
\affiliation{School of Mathematics, University of Bristol, University Walk, Bristol BS81TW, UK}
\affiliation{Toshiba Telecommuncations Research Laboratory, 32 Queens Square, Bristol BS1 4ND, United Kingdom}
\date{\today}

\begin{abstract}
We consider soft random geometric graphs, constructed by distributing points (nodes) randomly according to a Poisson Point Process, and
forming links between pairs of nodes with a probability that depends on their mutual distance, the ``connection function.''  Each
node has a probability of being isolated depending on the locations of the other nodes; we give analytic expressions for the
distribution of isolation probabilities.  Keeping the node locations fixed, the links break and reform over time, making a dynamic network;
this is a good model of a wireless ad-hoc network with communication channels undergoing rapid fading.
We use the above isolation probabilities to investigate the distribution of the time to transmit information to all the nodes, finding good
agreement with numerics. 
\end{abstract}

\maketitle

Many kinds of complex networks such as transport, power, social and
neuronal networks are spatial in character~\cite{Bart11}, that is, the nodes and perhaps also links have a physical location.
Geometry structures the network in that the probability of a link between two nodes is related to their mutual distance.

Consider a wireless ad-hoc network where nodes (devices) communicate directly with each other rather than a central router and where their locations
may be considered random;  examples include sensor~\cite{GT12} and vehicular~\cite{ZHCIH12} networks and the Internet of Things~\cite{RTBBA13}.
In wireless networks the probability of a
link decreases with the distance between nodes.  As time evolves, the links form a dynamic network~\cite{ZMN16}.
The communication channel exhibits rapid fading, so that some time later, the state of the system may be chosen independently with the same
distance-dependent probabilities.  Here we assume that the nodes remain in fixed locations, at least on the rapid fading timescale.  See supplemental
material~\cite{anim} for an animation in a square domain of length $L=8$ with $N=100$ nodes, showing connected components in different colours
and pausing when the whole network is connected.  The link probability between nodes of mutual distanace $r$ is Eq.~(\ref{e:Rayleigh}) below with
$\eta=2$ and $r_0=1$.

If the link probability is either zero or one everywhere, there is randomness only due to the node locations.  This is the case for the original random geometric graph
(RGG) model~\cite{Gilbert61}, in which nodes connect if and only if their mutual distance is less than a threshold $r_0$.  If the link probability somewhere lies strictly
between 0 and 1, there are two sources of randomness, in the node locations and the links.  Here, we fix the node locations (``quenched disorder''), and study the randomness
due to the links, as in the above dynamic wireless network application.  This system has also been studied using an approach based on graph entropy~\cite{Coon16}. 

We distribute nodes and links according to the following spatial inhomogeneous random graph model~\cite{Penrose15}: Place nodes in space
according to a Poisson Point Process (PPP) with intensity measure $\Lambda$ in $d$-dimensional space $\mathbb{R}^d$; we usually consider
$d\in\{1,2,3\}$.  This means that the number of nodes in a bounded set $A\subset\mathbb{R}^d$ is Poisson distributed with mean $\Lambda(A)$ and independent
of the number of points in any set $B$ disjoint with $A$.   Thus the average
number of nodes in the whole system is $\bar{N}=\Lambda(\mathbb{R}^d)$, possibly infinite.  The simplest case is where $\Lambda$ is proportional to Lebesgue measure,
that is, $\Lambda(A)=\rho {\rm Vol}(A)$ where $\rho$ is the (constant) density and ${\rm Vol}(A)$ is the volume of $A$.  In this case, we often replace $\mathbb{R}^d$ by a
cube $[0,L]^d$ with opposite faces identified (a flat torus).  Then $\bar{N}=\rho L^d$.

Now we form links between each pair of nodes with locations $\xi$, $\eta$, independently with probability $\phi(\xi,\eta)$.  Here we consider soft RGGs, for which
$\phi(\xi,\eta)=H(|\xi-\eta|)$ where $|.|$ denotes the Euclidean (or in general some other) length and $H:[0,\infty)\to[0,1]$
is called the connection function.  It is possible with information about node locations and links to construct a connection function for any
spatial network, and thus model it as a soft RGG. In practice, though,  the link independence assumption may not be accurate.  In the case of
wireless communication networks, there are detailed theories of the physics of the communication channel leading to a variety of connection functions; see
Refs.~\cite{DRW02,DG16,Penrose16}.

\begin{figure}
\centerline{\includegraphics[width=280pt]{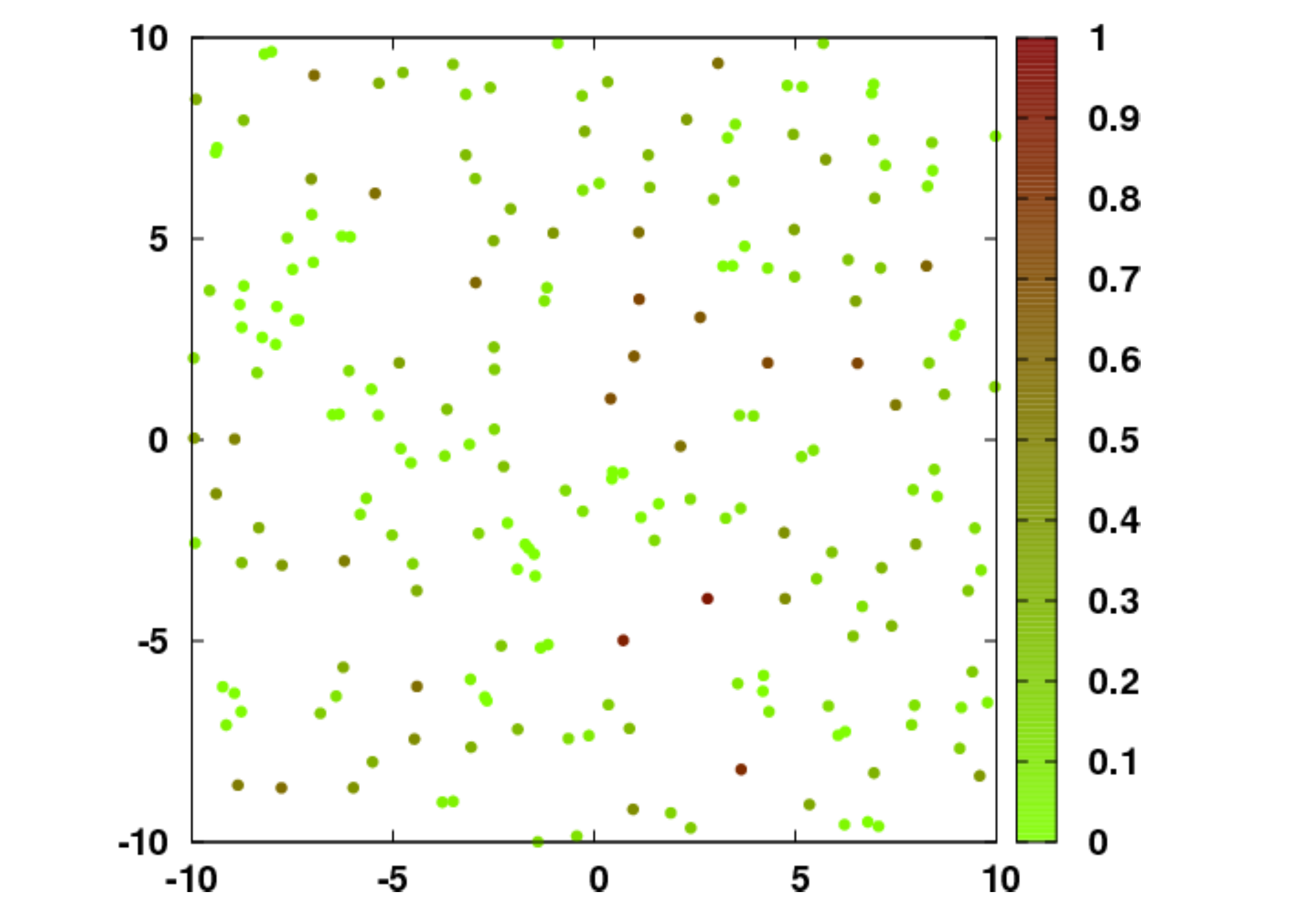}}
\caption{A Poisson Point Process with density $\rho=2$. Nodes are coloured by isolation probability, using the Rayleigh connection function, Eq.~(\protect\ref{e:Rayleigh})
with $\eta=2$ and periodic boundary conditions. \label{f:config}}
\end{figure}

Here, we use one of the simplest models: We assume Rayleigh fading, corresponding to diffuse scattering of the signal, which leads to an exponentially
distributed channel gain $|h|^2$.
The signal power decays as $r^{-\eta}$ where $\eta\in[2,6]$ is called the path loss exponent.  Free propagation gives the inverse square law
$\eta=2$, whilst more cluttered environments have a faster decay of the signal, leading to larger measured values of $\eta$.
A link may be made if the signal to noise ratio, proportional to $|h|^2r^{-\eta}$, reaches a given threshold, leading to the connection probability
\begin{equation}\label{e:Rayleigh}
H(r)=\exp\left(-(r/r_0)^\eta\right)
\end{equation}
for some constant $r_0$ that determines the length scale; we measure length in these units and so take $r_0=1$ hereafter.
Observe that the $\eta\to\infty$ limit gives the original RGG model.  

In order to understand transmission of information through a dynamic network, we must first analyse the instantaneous isolation probability of a
node, that is, the probability that it has no links.  This will determined by the locations of nearby nodes; see Fig.~\ref{f:config}.  Considering all nodes together, there is a
distribution of isolation probabilities.

The isolation probability of a node at $\xi$  in some  configuration $X$ of the PPP  is
\[ P_{iso}(\xi)=\prod_{\xi'\in X\backslash\{\xi\}}(1-H(|\xi'-\xi|)) \]
We note that in a PPP, the distribution of points found by conditioning on a node at a fixed position is unaffected (ie Palm
distribution of a PPP is the same PPP); see Ref.~\cite{Haenggi12} for the theory of PPPs.
To find the distribution of $P_{iso}$, we use the probability generating functional (PGF)
\[ G_X(u)=\mathbb{E}\prod_{\xi\in X}u(\xi)=\exp\left\{-\int(1-u(\xi))\Lambda(d\xi)\right\} \]
for arbitrary function $u(\xi)$ where the first equality is the definition and second follows for a PPP.  The function needs to satisfy some
mild conditions, for example (a) $\bar{N}<\infty$, or (b) $u\in[0,1]$ and $\int|\log u(\xi)|\Lambda(d\xi)<\infty$ as the case here.
In particular, \[ \mathbb{E}(P_{iso}(\xi))=\exp\left\{-\int H(|\xi'-\xi|) \Lambda(d\xi')\right\} \]
which is the connectivity mass, important for understanding the overall (multihop) connection probability of an ad-hoc network when
$d\geq 2$~\cite{DG16}.  However, we can also find further information about the distribution of $P_{iso}$,
namely for $\nu\in\mathbb{R}_{\geq 0}$ the $\nu$th moment is
\begin{equation}\label{e:Piso}
\mathbb{E}(P_{iso}(\xi)^\nu)=\exp\left\{-\int\left[1-(1-H(|\xi'-\xi|))^\nu\right]\Lambda(d\xi')\right\}
\end{equation}

If the PPP has constant density we find
\[ \mathbb{E}(P_{iso}^\nu)=\exp\left\{-S_d \rho\int_0^\infty\left[1-(1-H(r))^\nu\right]r^{d-1}dr\right\} \]
independent of the location of the node.  Here, $S_d$ is the total (solid) angle in $d$ dimensions, namely
$S_d=\{2,2\pi,4\pi\}$ for $d=\{1,2,3\}$ respectively.  Now, for Rayleigh fading, we have from Eq.~(\ref{e:Rayleigh}),
\begin{equation}\label{e:Pison}
\mathbb{E}(P_{iso}^\nu)=\exp\left[-\frac{S_d\rho}{\eta}\Gamma\left(\frac{d}{\eta}\right)H_\nu^{(d/\eta)}\right]
\end{equation}
where 
\begin{equation}\label{e:RHNi}
H_\nu^{(s)}=\frac{1}{\Gamma(s)}\int_0^\infty(1-(1-e^{-x})^\nu)x^{s-1}dx
\end{equation}
For integer $\nu=n$ we can expand the parentheses to yield a finite sum
\begin{equation}\label{e:RHNs}
H_n^{(s)}=\sum_{j=1}^n(-1)^{j-1}\left(\begin{array}{c}n\\j\end{array}\right)j^{-s}
\end{equation}
which is called the Roman harmonic number~\cite{Roman92}.

For $\nu$ not necessarily an integer,  we can use Eq.~(\ref{e:RHNi}): Numerical integration provides an efficient and stable means
of calculation, whilst asymptotically expanding the integral for large $\nu$ gives
\[ H_\nu^{(s)}=\frac{(\ln \nu)^s}{\Gamma(s+1)}+\frac{\gamma(\ln \nu)^{s-1}}{\Gamma(s)}+\frac{6\gamma^2+\pi^2}{12\Gamma(s-1)}(\ln \nu)^{s-2}+\ldots \]
where $\gamma\approx 0.5772$ is the Euler constant.  Thus we have
\begin{align}\nonumber
\mathbb{E}(P_{iso}^\nu)=&\exp\left[-V_d\rho\left((\ln \nu)^{\frac{d}{\eta}}+\frac{d}{\eta}\gamma(\ln \nu)^{\frac{d}{\eta}-1}\right.\right.\\
&\left.\left.+\frac{d}{\eta}\left(\frac{d}{\eta}-1\right)\frac{6\gamma^2+\pi^2}{12}(\ln \nu)^{\frac{d}{\eta}-2}+\ldots\right)\right]
\label{e:ntoinf}
\end{align}
where $V_d=S_d/d$ is the volume of the unit ball in $d$ dimensions. 

When $\eta=d$, that is, $s=1$, we have a further simplification
\begin{align}\label{e:digama}
\mathbb{E}(P_{iso}^\nu)&=\exp\{-V_d\rho (\gamma+\psi(\nu+1))\}\\
&=\exp\left\{-V_d\rho\left(\ln \nu+\gamma+\frac{1}{2\nu}-\frac{1}{12\nu^2}+\ldots\right)\right\}\nonumber
\end{align}
where $\psi(x)=\frac{\Gamma'(x)}{\Gamma(x)}$ is the digamma function, and we have used its standard expansion for large argument.  For integer
$\nu=n$
\[\gamma+\psi(n+1)= H_n=\sum_{j=1}^n\frac{1}{j} \]
is the usual harmonic number.

\begin{figure}
\centerline{\includegraphics[width=250pt]{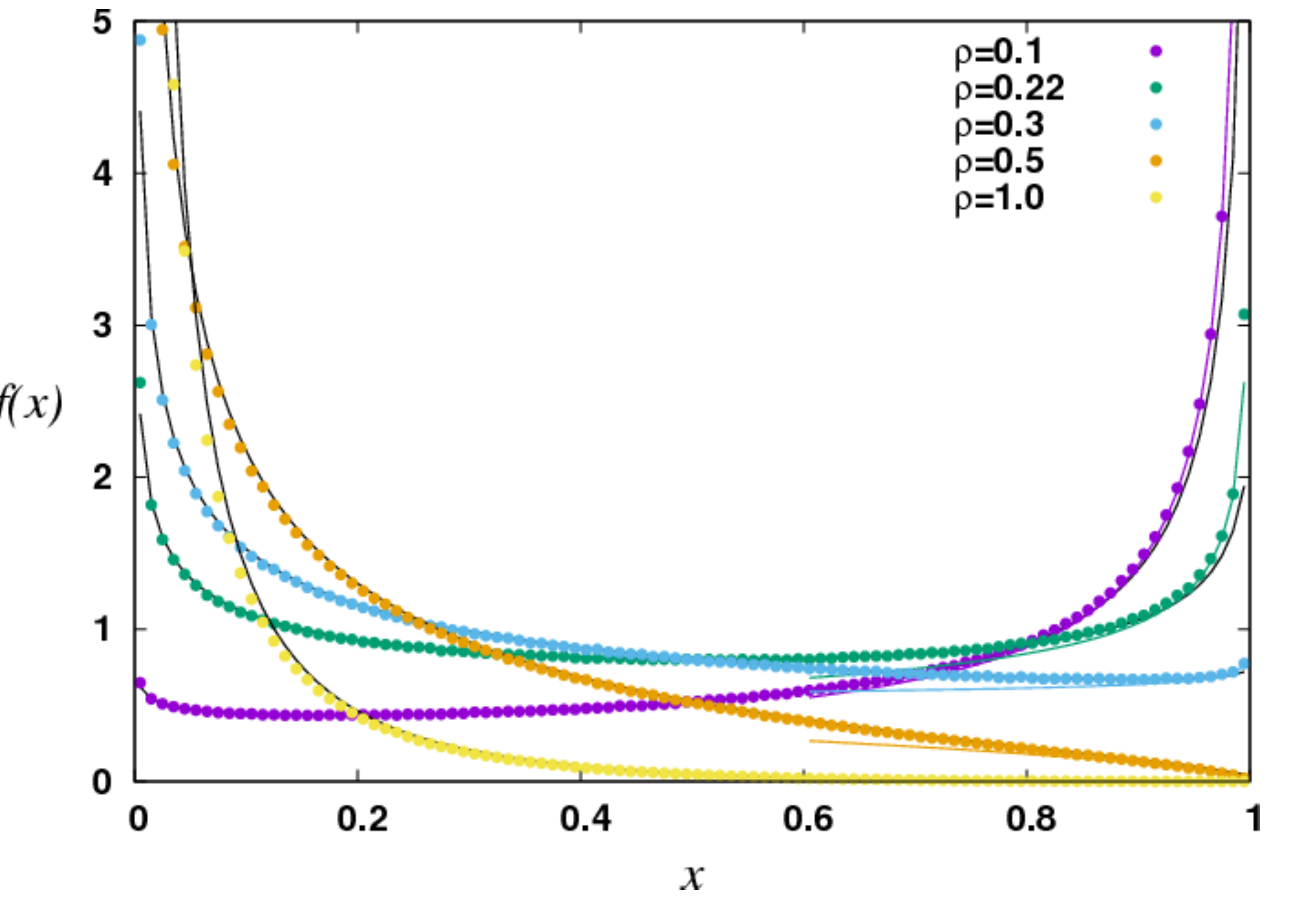}}
\centerline{\includegraphics[width=250pt]{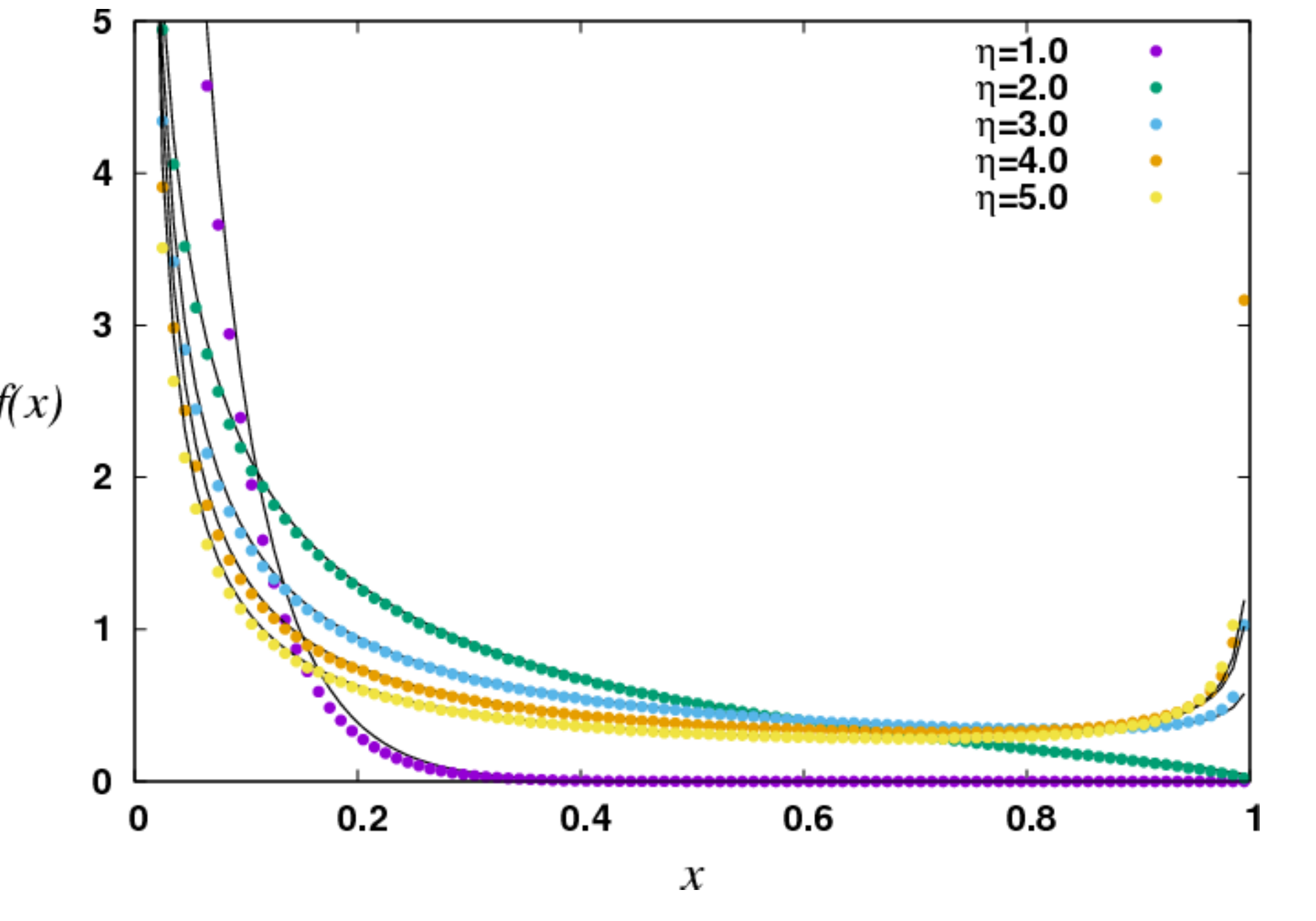}}
\caption{The pdf of $P_{iso}$, by direct simulation in $d=2$ (points) and using Eqs.~(\protect\ref{e:Pison},~\protect\ref{e:RHNs}~\protect\ref{e:mnat}) with $\alpha=100$ (black solid lines). Top: $\eta=2$.  Bottom: $\rho=0.5$.  For $\eta=2$ and $x>0.6$, the first term of Eq.~(\ref{e:asym}) is shown as a coloured line.\label{f:e2}}
\end{figure}

We can now attempt to extract the probability density function (pdf)
of $P_{iso}$, which we will denote as $f(x)$ with $x\in(0,1)$, from these moments.  For a general distribution on a finite interval
this is called the Hausdorff moment problem, and the solution is unique if it exists. For the most general numerical approach we follow
Mnatsakanov~\cite{Mnat08}, who gives for a general pdf determined from integer moments
$\mu_n=\mathbb{E}(P_{iso}^n)$, an approximation depending on a positive integer parameter $\alpha$:
\begin{equation}\label{e:mnat}
f_\alpha(x)=\frac{\Gamma(\alpha+2)}{\Gamma(a+1)}\sum_{m=0}^{\alpha-a}\frac{(-1)^m\mu_{m+a}}{m!(\alpha-a-m)!},\qquad a=\lfloor\alpha x\rfloor
\end{equation}
The function depends on $x$ only through $a$, and is hence piecewise constant for any fixed $\alpha$.  It converges to the correct function as $\alpha\to\infty$.
It is possible to use this to get a numerical approximation to $f(x)$, using high precision arithmetic to overcome problems from cancellations; see Fig.~\ref{f:e2}.
We see that $f(x)$ is singular at $x=0$ or $x=1$ or both; when $\eta=d=2$ it is almost symmetrical at $\rho=0.22$.  It is never quite symmetrical:
For $\mathbb{E}(P_{iso})=1/2$ we must have $\rho=\frac{\ln 2}{\pi}\approx 0.220636$ and then the third central moment is
$2^{-11/6}-3\times 2^{-5/2}+2^{-2}\approx 0.000285\neq 0$.

In the case of $\eta=d$, we can write
\[ \mathbb{E}(P^{\nu}_{iso})=\int_0^\infty f(x)x^{\nu} dx=\exp\{-V_d\rho(\psi(\nu+1)+\gamma)\} \]
giving a representation as an inverse Mellin transform
\[ f(x)=\frac{1}{2\pi i}\int_{c-i\infty}^{c+i\infty}x^{-s}\exp\{-V_d\rho(\psi(s)+\gamma)\}ds \]
This is however intractable either analytically or numerically.

Still for $\eta=d$, the large $\nu$ asymptotics does however give information on the behaviour of $f(x)$ near $x=1$, the distribution of highly isolated nodes.  Making an ansatz
\[ f(1-\epsilon)=\sum_{i=0}^\infty g_i \epsilon^{\delta+i} \]
multiplying by $(1-\epsilon)^\nu\approx \exp(-\nu\epsilon)$ and integrating gives
\[  \mathbb{E}(P^{\nu}_{iso})=\sum_{i=0}^\infty g_i\frac{\Gamma(\delta+i+1)}{\nu^{\delta+i+1}} \]
which by comparison with Eq.~(\ref{e:digama}) yields
\begin{equation}\label{e:asym}
f(1-\epsilon)=\epsilon^{V_d\rho-1}\frac{e^{-V_d\rho\gamma}}{\Gamma(V_d\rho)}\left[
1-\frac{\epsilon}{2}+\frac{\epsilon^2}{24}\frac{2+3V_d\rho}{1+V_d\rho}+\ldots\right]
\end{equation}
For moderate $\epsilon$ it is more accurate to keep just the first term, as shown in Fig.~\ref{f:e2}.

Alternatively, we can take the limit $d\to\infty$.  If $\eta$ increases proportional to $d$, we see from Eq.~(\ref{e:Pison}) that the only effect is to change the effective
density $\rho$.  If $\eta$ is constant, we find $H_{\nu}^{(d/\eta)}\approx\nu$, so that $\mathbb{E}(P_{iso}^\nu)\approx x_*^\nu$ with
\[ x_*=\exp\left[-\frac{S_d\rho}{\eta}\Gamma(\frac{d}{\eta})\right] \]
and so corresponds to a distribution that is sharply peaked at $x=x_*$.

Now, we return to the problem of a dynamic network, assuming $d>1$,  constant density and neglecting boundary effects.
The network chooses links anew each time $\tau$.  In order to ensure information can reach every node in the network, we need to ensure that
no node is isolated for the considered time interval $T\tau$ where $T$ is the number of time steps.  At each time step, the isolation probabilities
of the nodes are a PPP on $[0,1]$ with intensity $d\Lambda=\bar{N}f(x)dx$.  The probability that a node with $P_{iso}=x$ is isolated for $T$ consecutive timesteps
is simply $x^T$.  Denoting the event that none of the nodes are isolated during these $T$ timesteps by $C_T$ we can again use the PGF (noting that the number of
nodes is almost surely finite):
\[ \mathbb{P}(C_T|X)=\prod_i(1-x_i^T) \]
Averaging over configurations $X$ of the PPP,
\[ \mathbb{P}(C_T)=\exp\left\{-\bar{N}\int x^Tf(x)dx\right\} \]
But the integral is just $\mathbb{E}(P_{iso}^T)$ which we calculated in Eq.~(\ref{e:ntoinf}):
\begin{align}\nonumber
\mathbb{P}(C_T)
=&\exp\left\{-\rho L^d
\exp\left[-V_d\rho\left((\ln T)^{\frac{d}{\eta}}+\frac{d}{\eta}\gamma(\ln T)^{\frac{d}{\eta}-1}\right.\right.\right.\\
&\left.\left.\left.+\frac{d}{\eta}\left(\frac{d}{\eta}-1\right)\frac{6\gamma^2+\pi^2}{12}(\ln T)^{\frac{d}{\eta}-2}+\ldots\right)\right]\right\}
\label{e:NisoT}
\end{align}
from which we find that the time to ensure all nodes are connected at least once is
\[ T\approx \exp\left[\left(\frac{\ln (\rho L^d)}{V_d\rho}\right)^{\eta/d}\right] \]

\begin{figure}
\centerline{\includegraphics[width=250pt]{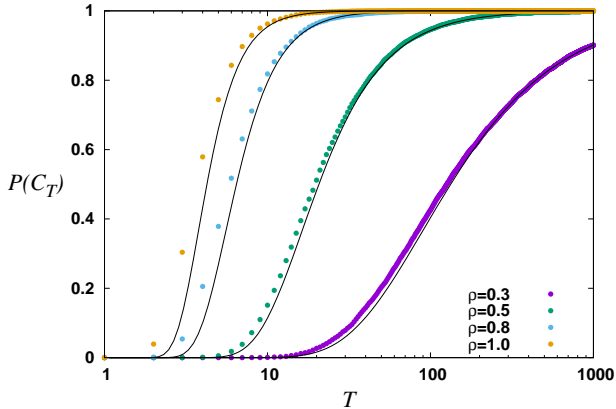}}
\caption{Probability that no nodes will remain isolated for time $T$.  Here, $d=\eta=2$, $L=20$, $10^4$ configurations: points.
Theoretical curves, Eq.~(\protect\ref{e:NisoT}).\label{f:NisoT}}
\end{figure}

Eq.~(\ref{e:NisoT}) has been confirmed by numerical simulation; see Fig.~\ref{f:NisoT}.  Thus, for low density the required time grows as a stretched
exponential, controlled by the path loss exponent $\eta$.  When $\eta=d$, it reduces to simply $T\approx(\rho L^d)^{1/(V_d\rho)}$, with the
probability distribution determined by the highly isolated nodes as in Eq.~(\ref{e:asym}).

Strictly speaking all our results for isolation probabilities apply to one dimensional networks.  However, in this case, transmission of information is limited by
large gaps, rather than nodes that are likely to be isolated.  In the original RGG, transmission can occur if and only if there are no gaps larger than the
link range $r_0$; see Ref.~\cite{Devroye81}.  For the soft RGG, it is quite likely that link may be made between
nodes that are not directly adjacent to the gaps, and estimating the probability of connectivity, even at a single point in time, remains an open problem. 

In conclusion, we have investigated the distribution of isolation probabilities in quenched soft random geometric graphs.  This has allowed an
analysis of the performance of a dynamic soft RGG model of wireless ad-hoc networks with fixed nodes and rapid channel fading. We obtained explicit
formulas for the probability that no node will be isolated for $T$ time steps, with good numerical agreement.
In contrast to networks with mobile nodes, the transmission of information is greatly hindered by extremes of the quenched
disorder, namely highly isolated nodes.  In the future it would be interesting to consider boundary effects, and other nonuniform
node distributions, which are characteristic of many spatial complex networks.

The authors would like to thank the directors of Toshiba Telecommunications Research Laboratory and the EPSRC (grant EP/N002458/1) for their support.
They are grateful to Justin Coon for helpful discussions.  
\bibliographystyle{apsrev4-1}
\bibliography{wireless}

%merlin.mbs apsrev4-1.bst 2010-07-25 4.21a (PWD, AO, DPC) hacked
%Control: key (0)
%Control: author (72) initials jnrlst
%Control: editor formatted (1) identically to author
%Control: production of article title (-1) disabled
%Control: page (0) single
%Control: year (1) truncated
%Control: production of eprint (0) enabled
\begin{thebibliography}{16}%
\makeatletter
\providecommand \@ifxundefined [1]{%
 \@ifx{#1\undefined}
}%
\providecommand \@ifnum [1]{%
 \ifnum #1\expandafter \@firstoftwo
 \else \expandafter \@secondoftwo
 \fi
}%
\providecommand \@ifx [1]{%
 \ifx #1\expandafter \@firstoftwo
 \else \expandafter \@secondoftwo
 \fi
}%
\providecommand \natexlab [1]{#1}%
\providecommand \enquote  [1]{``#1''}%
\providecommand \bibnamefont  [1]{#1}%
\providecommand \bibfnamefont [1]{#1}%
\providecommand \citenamefont [1]{#1}%
\providecommand \href@noop [0]{\@secondoftwo}%
\providecommand \href [0]{\begingroup \@sanitize@url \@href}%
\providecommand \@href[1]{\@@startlink{#1}\@@href}%
\providecommand \@@href[1]{\endgroup#1\@@endlink}%
\providecommand \@sanitize@url [0]{\catcode `\\12\catcode `\$12\catcode
  `\&12\catcode `\#12\catcode `\^12\catcode `\_12\catcode `\%12\relax}%
\providecommand \@@startlink[1]{}%
\providecommand \@@endlink[0]{}%
\providecommand \url  [0]{\begingroup\@sanitize@url \@url }%
\providecommand \@url [1]{\endgroup\@href {#1}{\urlprefix }}%
\providecommand \urlprefix  [0]{URL }%
\providecommand \Eprint [0]{\href }%
\providecommand \doibase [0]{http://dx.doi.org/}%
\providecommand \selectlanguage [0]{\@gobble}%
\providecommand \bibinfo  [0]{\@secondoftwo}%
\providecommand \bibfield  [0]{\@secondoftwo}%
\providecommand \translation [1]{[#1]}%
\providecommand \BibitemOpen [0]{}%
\providecommand \bibitemStop [0]{}%
\providecommand \bibitemNoStop [0]{.\EOS\space}%
\providecommand \EOS [0]{\spacefactor3000\relax}%
\providecommand \BibitemShut  [1]{\csname bibitem#1\endcsname}%
\let\auto@bib@innerbib\@empty
%</preamble>
\bibitem [{\citenamefont {Barth{\'e}lemy}(2011)}]{Bart11}%
  \BibitemOpen
  \bibfield  {author} {\bibinfo {author} {\bibfnamefont {M.}~\bibnamefont
  {Barth{\'e}lemy}},\ }\href@noop {} {\bibfield  {journal} {\bibinfo  {journal}
  {Physics Reports}\ }\textbf {\bibinfo {volume} {499}},\ \bibinfo {pages} {1}
  (\bibinfo {year} {2011})}\BibitemShut {NoStop}%
\bibitem [{\citenamefont {Goyal}\ and\ \citenamefont {Tripathy}(2012)}]{GT12}%
  \BibitemOpen
  \bibfield  {author} {\bibinfo {author} {\bibfnamefont {D.}~\bibnamefont
  {Goyal}}\ and\ \bibinfo {author} {\bibfnamefont {M.~R.}\ \bibnamefont
  {Tripathy}},\ }in\ \href@noop {} {\emph {\bibinfo {booktitle} {2012 Second
  International Conference on Advanced Computing \& Communication
  Technologies}}}\ (\bibinfo {organization} {IEEE},\ \bibinfo {year} {2012})\
  pp.\ \bibinfo {pages} {474--480}\BibitemShut {NoStop}%
\bibitem [{\citenamefont {Zeadally}\ \emph {et~al.}(2012)\citenamefont
  {Zeadally}, \citenamefont {Hunt}, \citenamefont {Chen}, \citenamefont
  {Irwin},\ and\ \citenamefont {Hassan}}]{ZHCIH12}%
  \BibitemOpen
  \bibfield  {author} {\bibinfo {author} {\bibfnamefont {S.}~\bibnamefont
  {Zeadally}}, \bibinfo {author} {\bibfnamefont {R.}~\bibnamefont {Hunt}},
  \bibinfo {author} {\bibfnamefont {Y.-S.}\ \bibnamefont {Chen}}, \bibinfo
  {author} {\bibfnamefont {A.}~\bibnamefont {Irwin}}, \ and\ \bibinfo {author}
  {\bibfnamefont {A.}~\bibnamefont {Hassan}},\ }\href@noop {} {\bibfield
  {journal} {\bibinfo  {journal} {Telecommunication Systems}\ }\textbf
  {\bibinfo {volume} {50}},\ \bibinfo {pages} {217} (\bibinfo {year}
  {2012})}\BibitemShut {NoStop}%
\bibitem [{\citenamefont {Reina}\ \emph {et~al.}(2013)\citenamefont {Reina},
  \citenamefont {Toral}, \citenamefont {Barrero}, \citenamefont {Bessis},\ and\
  \citenamefont {Asimakopoulou}}]{RTBBA13}%
  \BibitemOpen
  \bibfield  {author} {\bibinfo {author} {\bibfnamefont {D.~G.}\ \bibnamefont
  {Reina}}, \bibinfo {author} {\bibfnamefont {S.~L.}\ \bibnamefont {Toral}},
  \bibinfo {author} {\bibfnamefont {F.}~\bibnamefont {Barrero}}, \bibinfo
  {author} {\bibfnamefont {N.}~\bibnamefont {Bessis}}, \ and\ \bibinfo {author}
  {\bibfnamefont {E.}~\bibnamefont {Asimakopoulou}},\ }in\ \href@noop {} {\emph
  {\bibinfo {booktitle} {Internet of Things and Inter-Cooperative Computational
  Technologies for Collective Intelligence}}}\ (\bibinfo  {publisher}
  {Springer},\ \bibinfo {year} {2013})\ pp.\ \bibinfo {pages}
  {89--113}\BibitemShut {NoStop}%
\bibitem [{\citenamefont {Zhang}\ \emph {et~al.}()\citenamefont {Zhang},
  \citenamefont {Moore},\ and\ \citenamefont {Newman}}]{ZMN16}%
  \BibitemOpen
  \bibfield  {author} {\bibinfo {author} {\bibfnamefont {X.}~\bibnamefont
  {Zhang}}, \bibinfo {author} {\bibfnamefont {C.}~\bibnamefont {Moore}}, \ and\
  \bibinfo {author} {\bibfnamefont {M.}~\bibnamefont {Newman}},\ }\href@noop {}
  {\enquote {\bibinfo {title} {Random graph models for dynamic networks},}\
  }\bibinfo {howpublished} {arXiv:1607.07570}\BibitemShut {NoStop}%
\bibitem [{ani()}]{anim}%
  \BibitemOpen
  \href@noop {} {}\bibinfo {howpublished} {Supplemental material, file
  animated\_graph.gif}\BibitemShut {NoStop}%
\bibitem [{\citenamefont {Gilbert}(1961)}]{Gilbert61}%
  \BibitemOpen
  \bibfield  {author} {\bibinfo {author} {\bibfnamefont {E.~N.}\ \bibnamefont
  {Gilbert}},\ }\href@noop {} {\bibfield  {journal} {\bibinfo  {journal} {J.
  Soc. Indust. Appl. Math.}\ }\textbf {\bibinfo {volume} {9}},\ \bibinfo
  {pages} {533} (\bibinfo {year} {1961})}\BibitemShut {NoStop}%
\bibitem [{\citenamefont {Coon}(2016)}]{Coon16}%
  \BibitemOpen
  \bibfield  {author} {\bibinfo {author} {\bibfnamefont {J.~P.}\ \bibnamefont
  {Coon}},\ }in\ \href@noop {} {\emph {\bibinfo {booktitle} {IEEE Globecom 2016
  (accepted)}}}\ (\bibinfo {year} {2016})\BibitemShut {NoStop}%
\bibitem [{\citenamefont {Penrose}()}]{Penrose15}%
  \BibitemOpen
  \bibfield  {author} {\bibinfo {author} {\bibfnamefont {M.~D.}\ \bibnamefont
  {Penrose}},\ }\href@noop {} {\enquote {\bibinfo {title} {Inhomogeneous random
  graphs, isolated vertices, and {Poisson} approximation},}\ }\bibinfo
  {howpublished} {arXiv:1507.07132}\BibitemShut {NoStop}%
\bibitem [{\citenamefont {Durgin}\ \emph {et~al.}(2002)\citenamefont {Durgin},
  \citenamefont {Rappaport},\ and\ \citenamefont {De~Wolf}}]{DRW02}%
  \BibitemOpen
  \bibfield  {author} {\bibinfo {author} {\bibfnamefont {G.~D.}\ \bibnamefont
  {Durgin}}, \bibinfo {author} {\bibfnamefont {T.~S.}\ \bibnamefont
  {Rappaport}}, \ and\ \bibinfo {author} {\bibfnamefont {D.~A.}\ \bibnamefont
  {De~Wolf}},\ }\href@noop {} {\bibfield  {journal} {\bibinfo  {journal} {IEEE
  Transactions on Communications}\ }\textbf {\bibinfo {volume} {50}},\ \bibinfo
  {pages} {1005} (\bibinfo {year} {2002})}\BibitemShut {NoStop}%
\bibitem [{\citenamefont {Dettmann}\ and\ \citenamefont
  {Georgiou}(2016)}]{DG16}%
  \BibitemOpen
  \bibfield  {author} {\bibinfo {author} {\bibfnamefont {C.~P.}\ \bibnamefont
  {Dettmann}}\ and\ \bibinfo {author} {\bibfnamefont {O.}~\bibnamefont
  {Georgiou}},\ }\href@noop {} {\bibfield  {journal} {\bibinfo  {journal}
  {Physical Review E}\ }\textbf {\bibinfo {volume} {93}},\ \bibinfo {pages}
  {032313} (\bibinfo {year} {2016})}\BibitemShut {NoStop}%
\bibitem [{\citenamefont {Penrose}\ \emph {et~al.}(2016)\citenamefont {Penrose}
  \emph {et~al.}}]{Penrose16}%
  \BibitemOpen
  \bibfield  {author} {\bibinfo {author} {\bibfnamefont {M.~D.}\ \bibnamefont
  {Penrose}} \emph {et~al.},\ }\href@noop {} {\bibfield  {journal} {\bibinfo
  {journal} {The Annals of Applied Probability}\ }\textbf {\bibinfo {volume}
  {26}},\ \bibinfo {pages} {986} (\bibinfo {year} {2016})}\BibitemShut
  {NoStop}%
\bibitem [{\citenamefont {Haenggi}(2012)}]{Haenggi12}%
  \BibitemOpen
  \bibfield  {author} {\bibinfo {author} {\bibfnamefont {M.}~\bibnamefont
  {Haenggi}},\ }\href@noop {} {\emph {\bibinfo {title} {Stochastic geometry for
  wireless networks}}}\ (\bibinfo  {publisher} {Cambridge University Press},\
  \bibinfo {year} {2012})\BibitemShut {NoStop}%
\bibitem [{\citenamefont {Roman}(1992)}]{Roman92}%
  \BibitemOpen
  \bibfield  {author} {\bibinfo {author} {\bibfnamefont {S.}~\bibnamefont
  {Roman}},\ }\href@noop {} {\bibfield  {journal} {\bibinfo  {journal} {The
  American mathematical monthly}\ }\textbf {\bibinfo {volume} {99}},\ \bibinfo
  {pages} {641} (\bibinfo {year} {1992})}\BibitemShut {NoStop}%
\bibitem [{\citenamefont {Mnatsakanov}(2008)}]{Mnat08}%
  \BibitemOpen
  \bibfield  {author} {\bibinfo {author} {\bibfnamefont {R.~M.}\ \bibnamefont
  {Mnatsakanov}},\ }\href@noop {} {\bibfield  {journal} {\bibinfo  {journal}
  {Statistics \& Probability Letters}\ }\textbf {\bibinfo {volume} {78}},\
  \bibinfo {pages} {1869} (\bibinfo {year} {2008})}\BibitemShut {NoStop}%
\bibitem [{\citenamefont {Devroye}(1981)}]{Devroye81}%
  \BibitemOpen
  \bibfield  {author} {\bibinfo {author} {\bibfnamefont {L.}~\bibnamefont
  {Devroye}},\ }\href@noop {} {\bibfield  {journal} {\bibinfo  {journal} {The
  Annals of Probability}\ }\textbf {\bibinfo {volume} {9}},\ \bibinfo {pages}
  {860} (\bibinfo {year} {1981})}\BibitemShut {NoStop}%
\end{thebibliography}%

\end{document}